\documentclass[manuscript]{aastex}
\begin{document}
\title{13 Color Photometry of Open Cluster M48}
\author{Zhen-Yu Wu, Xu Zhou, Jun Ma, Zhao-Ji Jiang, Jian-Sheng Chen}
\affil{National Astronomical Observatories, Chinese Academy of Sciences, 20A
Datun Road, Beijing 100012, China} \email{zywu@bac.pku.edu.cn}
\begin{abstract}
This paper presents 13 color CCD intermediate-band spectrophotometry of a field
centered on the open cluster M48 (NGC 2548), from 400nm to nearly 1000nm, taken
with Beijing-Arizona-Taiwan-Connecticut (BATC) Multi-Color Survey Photometric
system. The fundamental parameters of this cluster are derived with a new
method which based on the comparison between the spectral energy distributions
(SEDs) of cluster stars and the theoretical SEDs of Padova models. We find that
the best-fitting age of M48 is 0.32 Gyr, the distance is 780 pc and a reddening
$E(B-V)=0.04$ with a solar metallicity $Z=0.019$.
\end{abstract}
\section{Introduction}
Open clusters have long been recognized as important tools in the study of the
Galactic disk. Their value lies in the improvement in accuracy for distance
determination, metal content, reddening and age produced by the collective
stellar sample which shares these properties. So open clusters are excellent
tracers of the abundance gradient along the Galactic disk as well as many other
important disk properties, such as the age-metallicity relation, abundance
gradient evolution, disk age, and so on \citep{fr95, tw97, ch03}.

 The open cluster M48, also known as NGC 2548 is a quite conspicuous object
and should be a naked-eye object under good weather conditions. It was firstly
studied by \citet{pesch} who made $UBV$ photoelectric photometry of 37 stars,
from which the cluster turned out to have a reddening $E(B-V) = 0.04 \pm 0.05$
and a distance 630 pc. DDO photoelectric photometry of 5 red giants was
obtained by \citet{claria}. From 4 members of these red giants, he conclude
that this cluster has a metallicity [Fe/H]=0.1, $E(B-V) = 0.06 \pm 0.02$ and a
distance of 530 pc. \citet{merm} derived an age of 0.30 Gyr based on a
synthetic composite color-magnitude diagram. \citet{wu} determined absolute
proper motions and membership probabilities for 501 stars in the field of M48.
More recently, \citet{rid03} obtained a new result of M48 taken in the
$u^{'}g^{'}r^{'}i^{'}z^{'}$ SDSS filter system. They find that a distance of
700 pc, an age of 0.40 Gyr and a metallicity of [Fe/H]=0.0 can fit their data
best.

In this paper we present a new photometric result of M48 taken with BATC
Multi-Color Survey Photometric system. The BATC filter system consists of 15
filters of band-widths 150 -- 350\r{A} that cover the wavelength range
3300\r{A} -- 10000\r{A}, which avoids strong and variable sky emission lines
\citep{fan}. As the first object in BATC survey, the old open cluster M67 has
been studied based on color-magnitude diagram (CMD)\citep{fan}. Using the BATC
filter system, \citet{ch01} studied the globular cluster NGC 288 by comparing
SEDs of bright stars with Kurucz models. The estimated effective temperatures
and average value of [Fe/H] for these stars are consistent with spectroscopic
determinations. Based on the BATC survey observations, the main aim of this
study is to determine simultaneously the fundamental parameters of M48, such as
age, distance, metallicity and reddening, by comparing observational SEDs of
cluster stars with theoretical stellar evolutionary models.

The observations and reduction of the M48 data are described in Sec. 2. In Sec.
3, we derive fundamental parameters of M48. Conclusions and summary are
presented in Sec. 4.
\section{Observation and Data Reduction}
\subsection{Observation}
The observations are done with BATC photometric system at Xinglong Station of
National Astronomical Observatories, Chinese Academy of Sciences (NAOC). The
60/90 cm f/3 Schmidt telescope is used with a Ford Aerospace $2048\times2048$
CCD camera at its main focus. The field of view of the CCD is
$58\arcmin\times58\arcmin$ with a plate scale of 1.7 arcsec per pixel.

The filter system of BATC  project is defined by 15 intermediate-band filters
which are designed specifically to avoid most of the known bright and variable
night sky emission lines. The definition of magnitude for the BATC survey is in
the $AB_{\nu}$ system which is a monochromatic $F_{\nu}$ system first
introduced by \citet{og}:
\begin{equation}
m_{\textrm{batc}}=-2.5\,\log\,{F_{\nu}}-48.60
\end{equation}
where $F_{\nu}$ is the appropriately averaged monochromatic flux (measured in
unit of $\mbox{erg}~\mbox{s}^{-1}~\mbox{cm}^{-2}~\mbox{Hz}^{-1}$) at the
effective wavelength of the specific pass-band \citep{fuku}. In BATC system,
the $F_{\nu}$ is defined as \citep{yan}:
\begin{equation}
F_{\nu}= \frac{\int\textrm{d}(\log \nu)f_{\nu}R_{\nu}}{\int\textrm{d}(\log
\nu)R_{\nu}}
\end{equation}
which ties directly the magnitude to input flux. The system response $R_{\nu}$
actually used to relate the spectrum energy distribution of the source
$f_{\nu}$ and $F_{\nu}$, includes only the filter transmissions. Other effects,
such as the quantum efficiency of the CCD, the response of the telescope's
optics, the transmission of atmosphere, etc., are ignored. This makes the BATC
system filter-defined, and the effective wavelengths are affected only at the
$\pm 6$\r{A} level after taking CCD quantum efficiency and aluminum reflection
into account \citep{yan}.

The flux calibration of the BATC photometric system is defined by four
spectrophotometric standard stars of \citet{og}: HD 19445, HD 84937,
BD+26\degr2606 and BD+ 17\degr4708. The fluxes of the 4 standard stars have
been re-calibrated by \citet{fuku}. Their magnitudes in BATC system have been
slightly corrected recently by cross checking with the data obtained on a
number of photometric nights \citep{zhou01}.

In the nights judged photometric by the observers, the standard stars were
observed between air-masses 1.0 and 2.0 for each filter band. The observing
procedure of survey program field and photometry are described in detail in
\citet{zhou01} and \citet{yan}.

Because of the very low quantum efficiency of the thick CCD used in the bluest
filters, two BATC filters $a$ and $b$ are not used in the observation of M48
field.  In Table \ref{tb1}, for each BATC filter, we list the corresponding
effective wavelength, FWHM and exposure time.

\subsection{Data Reduction}
Preliminary reductions of  the CCD frames, including bias subtraction and field
flattening, were carried out with an automatic data reduction procedure called
\emph{Pipline I} which has been developed as a standard for the BATC survey in
NAOC \citep{fan}. The astrometric plate solution is obtained by \emph{a priori}
knowing the approximate plate center position, and then using this information
to register the brighter stars in each frame with the Guide Star Catalog (GSC)
coordinate system \citep{jen}.

A \emph{Pipeline II} program which based on the DAOPHOT II stellar photometric
reduction package of Stetson \citep{stet} was used to measure the instrumental
magnitudes of point sources in BATC CCD frames. The \emph{Pipeline II}
reduction procedure was performed on each single CCD frame to get the PSF
magnitude of each point source. The instrumental magnitudes were then
calibrated to BATC standard system \citep{zhou03}. The average calibration
error of each filter is less than 0.02 mag. The other sources of photometric
error include photon statistics of star and sky, readout noise, random and
systematic errors from bias subtraction and flat-fielding, and the PSF fitting,
are considered in \emph{Pipeline II} and the total estimated errors of each
star are given in the final catalog \citep{zhou03}. Stars which are detected in
at least 3 filters are included in the final catalog.

\section{Fundamental Parameters Derived from SEDs}
CMD is the main tool to derive the fundamental parameters of star cluster. In
the BATC photometric system, 15 filters can be used to form various CMDs.
However, it is difficult to derive a consistent result from various different
CMDs. In the other hand, the large number of bands available in our data set
provides a sort of \emph{low resolution spectroscopy}, which defines the SED of
each star quite well. So it is possible to derive the fundamental parameters of
star cluster by fitting the SEDs of member stars  with theoretical stellar
evolutionary models.
\subsection{Fitting Procedure and Results}
The observed SED of a star is determined by intrinsic (mass, age, metallicity)
and extrinsic (distance and reddening) parameters. The member stars in a star
clusters share the same age, metallicity, reddening and distance. The idea of
our new method is to compare the observed SEDs of member stars with theoretical
models to obtain a combination of best-fitting cluster parameters.

Our fitting procedure can be separated into two steps. First, we calculate the
deviation between observed and theoretical SEDs of each member star with
different sets of cluster parameters including age, metallicity, distance and
reddening:
\begin{equation}
m_{x}[E(B-V), r]=m_{x_{obs}}+5-5\log r-R_{x}\times E(B-V)
\end{equation}
where $m_{x_{obs}}$ is the observed magnitude of a cluster member in filter
$x$, $m_{x}[E(B-V),r]$ is the absolute magnitude corrected by distance $r$ and
reddening $E(B-V)$, $R_{x}$ is extinction coefficient which transform the
$E(B-V)$ to the BATC filter system and derived by \citet{ch00} based on the
procedure in Appendix B of \citet{sch}.

The criteria of fit for a cluster member $s$ between observed SED and
theoretical model is defined by:
\begin{equation}
\zeta_{s}\,[\log(t),Z,m,r,E(B-V)]=\sqrt{\frac{\sum\{m_{x}[E(B-V),r]-M_{x}\}^{2}}{n}}
\end{equation}
where $M_{x}$ is the theoretical magnitude of a star in filter $x$, computed
from an chosen theoretical isochrone models with age $\log(t)$, metallicity $Z$
and mass $m$, $n$ is the number of filters used in the fit.

For any set of cluster parameters combination, we can find the smallest value
of $\zeta_{s}$ viz. $MIN_{\zeta_{s}\,[\log(t),Z,r,E(B-V)]}$ for the member star
$s$ and give the best-matched theoretical mass of that star. Then we can get
the total deviation of all member stars of cluster under this set of
fundamental parameters.
\begin{equation}
\zeta\,[\log(t),Z,r,E(B-V)]=\frac{\sum{MIN_{\zeta_{s}\,[\log(t),Z,r,E(B-V)]}}}{N}
\end{equation}
where $N$ is the number of cluster members used in the fit. For a set of
fundamental parameters that best match the observed and theoretical SEDs of
cluster members will yield the minimum of $\zeta$.

As our request, Dr.~L.~Girardi has kindly calculated isochrones of our filter
system using the known BATC filter transmission curves and their Padova stellar
evolutionary models \citep{gir00,gir02}. The Padova isochrone sets are computed
with updated opacities and equations of state, and a moderate amount of
convective overshoot.

The results of \citet{wu} proper motion and membership study of M48 are used to
determine members in this cluster. Stars with membership probabilities greater
than 0.7 are considered to be members \citep{wu}. All stars considered as
members based on their proper motions were used in our fitting.

In our fitting procedure, the distances was chosen from 600 pc to 900 pc in
intervals of 10 pc, $E(B-V)$ from 0.00 to 0.10 in intervals of 0.01.
Theoretical isochrone models with metallicity $Z=$ 0.08, 0.019, 0.030 and age
$\log(t)$ from 8.0 to 9.0 in intervals of 0.05 were chosen.

\begin{figure}
\plotone{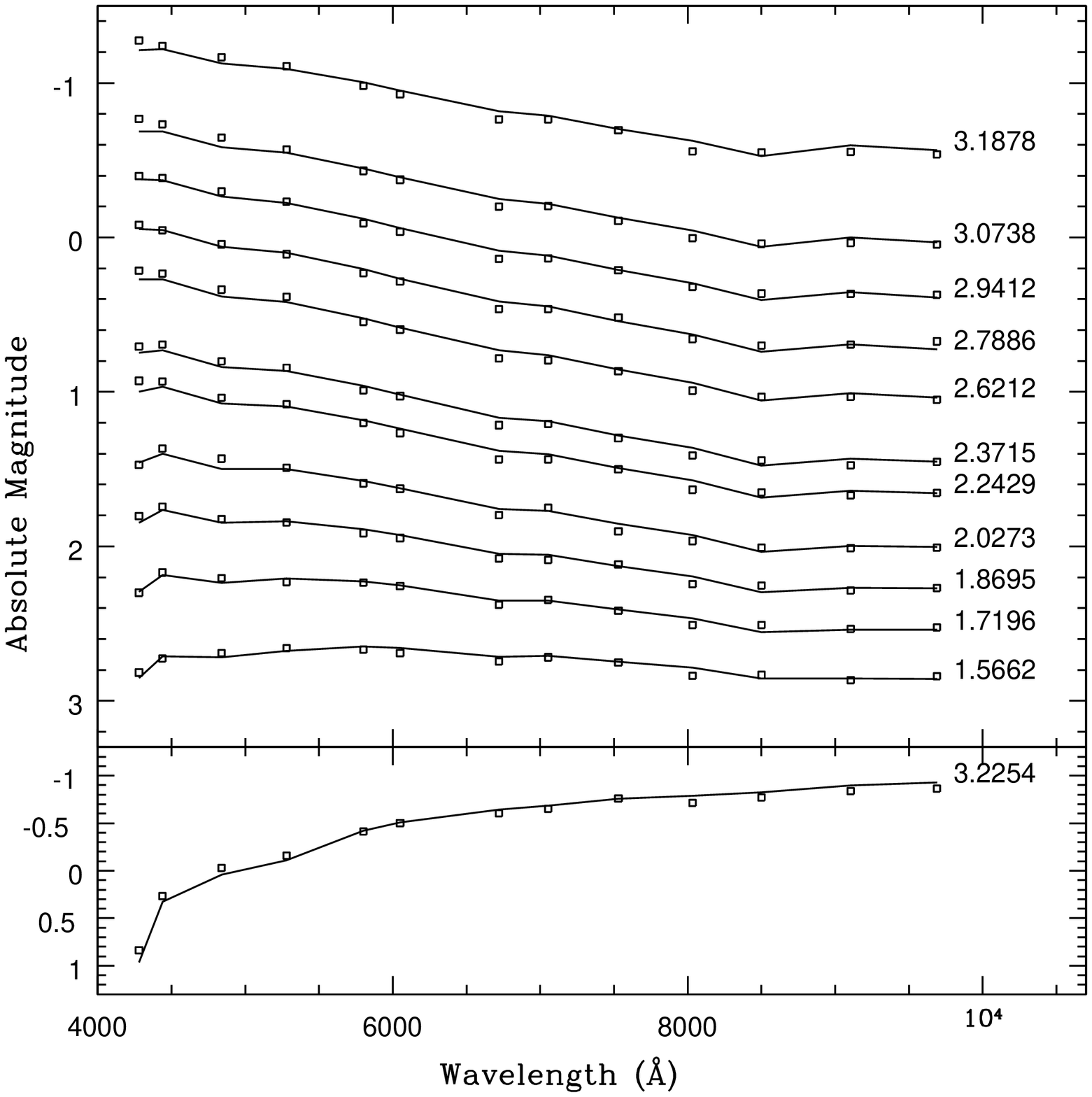}\caption{SEDs of M48 members of BATC data compared with the
Padova models. The open squares are the observations plotted at the effective
wavelengths of the BATC filters. The line is the best-fitting theoretical SED
from \citet{gir00}. For each line, the corresponding mass is labelled on the
right in the unit of solar mass $M_\sun$. \label{sed}}
\end{figure}

We find that, with a distance of 780 pc, reddening $E(B-V)=0.04$, the
theoretical model with age of 0.32 Gyr and metallicity $Z=0.019$ give the
smallest value of $\zeta$ and best fit the observed SEDs. In Figure \ref{sed},
we plot the best-fitting results for some member stars. The mass of each object
in Padova models is labelled on the right of each corresponding curve in the
unit of solar mass $M_\sun$. In the top panel of Figure \ref{sed}, SEDs of 11
main sequence (MS) stars with mass from 1.5662 $M_\sun$ to 3.1878 $M_\sun$ are
plotted. In the bottom panel of Figure \ref{sed}, the SED of a red giant star
is plotted. We can see that the observed SEDs of both MS stars and red giant
stars with the derived best-fitting parameters, can fit the theoretical ones
very well.

\begin{figure}
\plotone{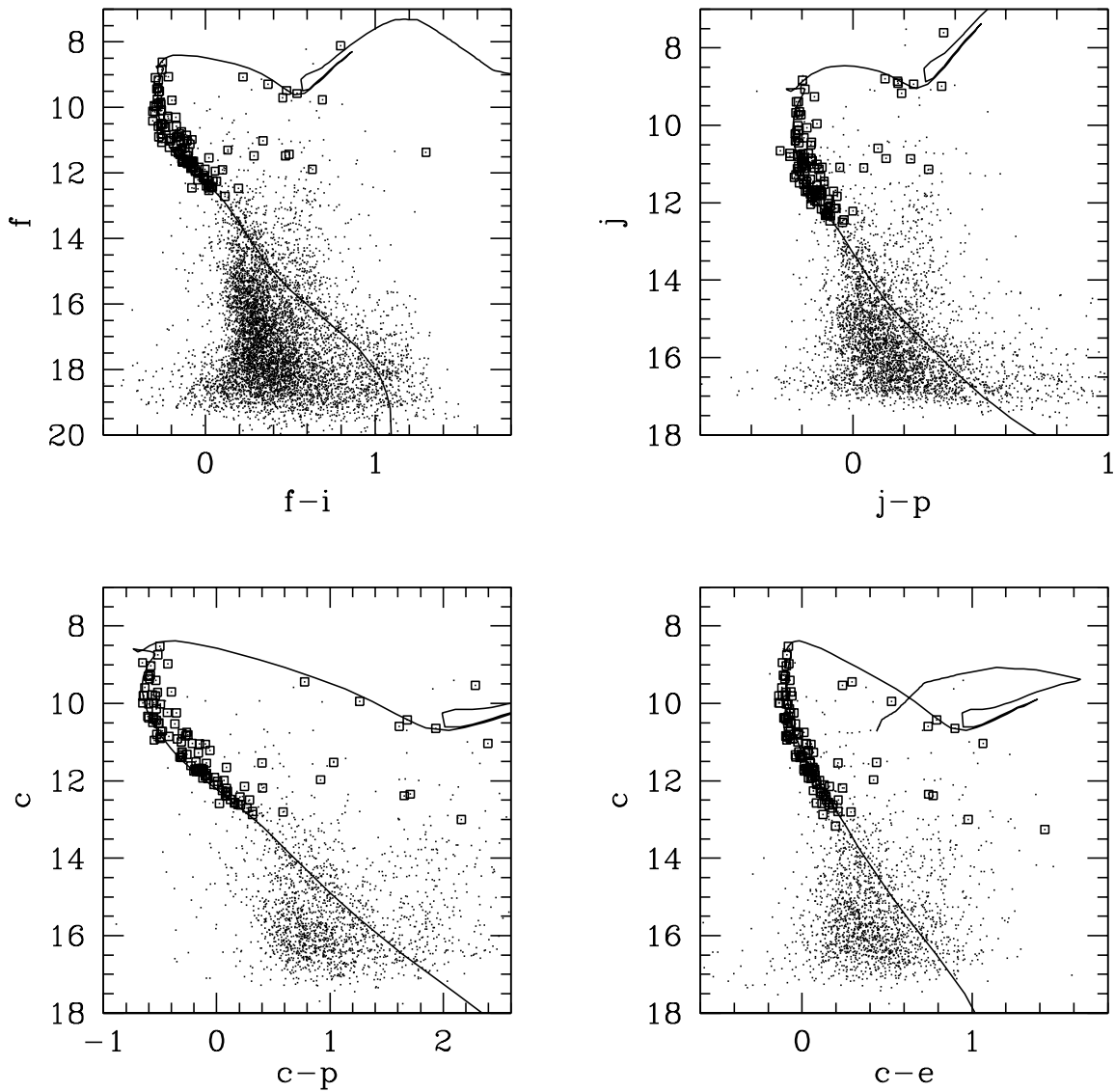} \caption{The color-magnitude diagrams for M48. Stars with
membership probabilities greater than 0.7 considered as members \citep{wu} are
plotted as squares. A distance of 780 pc, $E(B-V)=0.04$ and a metallicity of
$Z=0.019$ are adopted, a Padova isochrone with age of 0.32 Gyr is over-plotted
in each CMD. \label{cmd}}
\end{figure}

In Figure \ref{cmd}, we plot four representative CMDs from our data: ($c-p$) vs
$c$ gives us the widest pass-band colors; ($c-e$) vs $c$ gives us the cleanest
CMD; and ($f-i$) vs $f$ give us the deepest CMD. All of the CMDs have a well
defined MS and MS turnoff point. All stars in the field of M48 are plotted in
each diagram. Stars with known membership probabilities greater than 0.7
determined from the proper motion study of \citet{wu}, are given distinct
plotting symbols. In each CMD of Figure \ref{cmd}, parameters derived from
SED-fitting are adopted. The theoretical isochrone with the parameters derived
by SED-fitting can fit star's distribution in each CMD very well.

The uncertainties in our derived best-fitting fundamental parameters can be
determined from the observational photometric errors. For each member of
cluster that used in previous fitting procedure, a new magnitude in each BATC
filter was generated in a Monte Carlo fashion by adding Gaussian deviates to
the observed magnitude. The standard deviations of the deviates in each BATC
filter were taken to be 0.05 mag which is the maximum photometric error in most
filters of used sample stars. All of these stars with new artificial magnitudes
were then fitted by the same procedure as previous section. We repeated this
simulation 100 times. In the end, we find that the uncertainty caused by
photometric errors in distance is $\pm10$ pc, in reddening $E(B-V)$ is
$\pm0.01$, in age $\log(t)$ is $\pm0.05$. The obtained metallicity keeps the
same value. So the effect of photometric errors on the best-fitting results is
very small. At the same time, there still remains the possibility that there
could be some sort of systematic calibration problem between the photometry and
the models, or systematic errors in this particular set of photometry, although
such errors are small, and not necessarily zero.

In Table \ref{tb2} we list the fundamental parameters of M48 derived from this
and previous works. The derived age of this work is consistent with those
derived by previous works, the largest difference is 0.08 Gyr \citep{rid03}.
Our derived distance is very close to these results derived in resent years
\citep{dias02, rid03}. Our derived metellicity is consistent with that of
\citet{rid03} and close to the value derived by \citet{claria}. The reddening
$E(B-V)$ derived in this work is same as that of \citet{pesch} and very close
to those derived by previous works.

\section{Conclusions}
In this paper, we present and discuss new BATC multicolor photometry results
for the intermediate age open cluster, M48. Comparing the observed SEDs of
cluster member stars with the theoretical SEDs of Padova models, we find a set
best-fitting fundamental parameters for this cluster: an age of 0.32 Gyr, a
distance of 780 pc, a metallicity of $Z=0.019$ with a reddening $E(B-V)=0.04$.
This SED-fitting result can also fit CMDs formed from our data very well. Our
derived values are also consistent with those derived by previous authors
\citep{merm,rid03}. So, we can say that using the Padova theoretical
isochrones, we can effectively fit our data obtained based on BATC filter
system with theoretical isochrones and SEDs to extract useful cluster
information.

\acknowledgments

We would like to thank the anonymous referee for his/her insightful comments
and suggestions that improved this paper. We wish to thank Dr. Leo Girardi for
his assistance in the theoretical models. Z.Y.W acknowledge Dr. Wen-Ping Chen
for many useful suggestions. This work has been supported by the Chinese
National Key Basic Research Science Foundation (NKBRSF TG199075402), in part by
the Chinese National Science Foundation, No. 10473012 and 10373020.


\clearpage
\begin{deluxetable}{ccccc}
\tablenum{1} \tablewidth{0pt} \tablecaption{Parameters of the 13 BATC filters
and statistics of observations \label{tb1}} \tablehead{\colhead{No.} &
\colhead{Filter} & \colhead{$\lambda_{\textrm{eff}}$\, (\AA) }&\colhead{FWHM
(\AA)} &\colhead{Exposure time (min)}} \tablecolumns{5} \startdata
1&$c$&4180.0&310.0&38\\
2&$d$&4532.0&330.0&183\\
3&$e$&4916.0&370.0&98\\
4&$f$&5258.0&340.0&84\\
5&$g$&5785.0&290.0&64\\
6&$h$&6069.0&310.0&41\\
7&$i$&6646.0&490.0&44\\
8&$j$&7055.0&240.0&11\\
9&$k$&7545.0&190.0&21\\
10&$m$&8020.0&260.0&76\\
11&$n$&8483.0&170.0&88\\
12&$o$&9180.0&250.0&148\\
13&$p$&9736.0&280.0&134\\
\enddata
\end{deluxetable}

\begin{deluxetable}{ccccc}
\tablenum{2} \tablewidth{0pt} \tablecaption{Fundamental parameters of
M48\label{tb2}}\tablehead{\colhead{Author}&\colhead{Distance} & \colhead{Age} &
\colhead{Reddening}&\colhead{Metallicity}\\
\colhead{}&\colhead{pc}&\colhead{Gyr}&\colhead{$E(B-V)$}&\colhead{[Fe/H]}}
\tablecolumns{5} \startdata
 This work&780&0.32&0.04&0.0\\
 \citet{rid03}&700&0.40&0.03&0.0\\
 \citet{dias02}&770&0.36&0.03&0.08\\
 \citet{claria}&530&&0.06&0.1\\
 \citet{pesch}&630&&0.04&\\
 \citet{merm}&&0.30&&\\
 \enddata
\end{deluxetable}
\end{document}